\documentclass{moriond}

\usepackage{amsmath}
\renewcommand{\vec}[1]{{\mathbf #1}}

\newcommand{\ep}{\epsilon}

\newcommand{\bt}{\beta}
\newcommand{\g}{\gamma}

\newcommand{\dt}{\delta}

\newcommand{\Or}{\mathcal O}

\newcommand{\vL}{\ensuremath{\mathcal{L}}}

\newcommand{\sq}{^{2}}

\newcommand{\MS}{$\overline{\rm MS}$ }

\newcommand{\dslash}[1]{#1 \llap{/\kern-0.5pt}}
\newcommand{\Dslash}[1]{#1 \llap{/\kern+1.5pt}}
\newcommand{\DDslash}[1]{#1 \llap{/\kern+2.3pt}}
\newcommand{\dslashh}[1]{#1 \llap{/\kern+1pt}}

\newcommand{\boldsigma}{\mbox{\boldmath $\sigma$}}

\newcommand{\bea}{\begin{eqnarray}}
\newcommand{\eea}{\end{eqnarray}}
\newcommand{\be}{\begin{equation}}
\newcommand{\ee}{\end{equation}}
\newcommand{\bma}{\begin{pmatrix}}
	\newcommand{\ema}{\end{pmatrix}}
\newcommand{\nn}{\nonumber}

\newcommand{\NLDBD}{$0 \nu \beta \beta$}

\newcommand{\nnpp}{$nn \rightarrow p p\, e^- e^-$ }

\newcommand{\Photo}{}

\begin{document}
\vspace*{0cm}
\title{Neutrinoless double beta decay in effective field theory}

\author{ W. Dekens }

\address{Department of Physics, University of California at San Diego, 9500 Gilman Drive,\\ La Jolla, CA 92093-0319, USA
}

\maketitle\abstracts{
We discuss the contributions of lepton-number-violating (LNV) sources to neutrinoless double beta decay (\NLDBD). Assuming that these sources arise at scales well above the electroweak scale, they can be described within an effective field theory. Here, we outline the steps required to express the \NLDBD\ half-life in terms of the effective interactions, focusing on the dimension-five operator that induces a Majorana mass for the neutrinos.  This process involves the evolution of the operators down to scales of a few GeV where they can be matched onto Chiral Perturbation Theory.  The resulting Chiral Lagrangian can then used be to derive the lepton-number violating potential, which, in combination with many-body methods, gives the \NLDBD\ half-life.  
We will show that consistent renormalization requires the inclusion of a new contact interaction at leading order in this potential. We also briefly comment on the constraints that can be set on the operators appearing beyond dimension five.
}

\section{Introduction}
The observation of neutrinoless double beta decay would imply that neutrinos are Majorana particles~\cite{Schechter:1981bd}, show that lepton number is violated,  and signal physics beyond the Standard Model (SM). Current experiments already set stringent limits on the half-life of this process, e.g.\ \cite{KamLAND-Zen:2016pfg} $T_{1/2}^{0\nu}>1.07\cdot 10^{26}\,y$ in $^{136}$Xe, while  next-generation experiments aim for one to two orders of magnitude improvement in sensitivity.

Perhaps the most familiar contributions to \NLDBD\ are due to the exchange of light Majorana neutrinos, however, various beyond-the-SM (BSM) scenarios give rise to different types of LNV sources. If one assumes that this LNV is induced at a scale $\Lambda$ well above the electroweak scale, these sources can be described in an effective field theory (EFT), the so-called SM-EFT. Within this EFT, BSM effects are parametrized by higher-dimensional operators which are suppressed by powers of the high scale, $\Lambda$. 
The information about a particular BSM scenario is then captured by the coefficients of the higher-dimensional operators. Thus, after expressing the \NLDBD\ half-life in terms of these coefficients, the task of assessing the impact of a particular BSM scenario is reduced to simply matching it to the EFT.

The  steps to derive the contributions of the effective operators to \NLDBD\ involve the evolution of the interactions to the electroweak scale, where the heavy SM fields are integrated out. The EFT can then be evolved to scales of a few GeV where QCD becomes non-perturbative. At this point one matches the quark-level theory onto Chiral EFT, where the degrees of freedom are nucleons, pions, and leptons. 
The Chiral interactions come with unknown low-energy constants (LECs), so that one relies on a power-counting scheme to determine their relative importance.
We will show that Weinberg's power-counting scheme~\cite{Weinberg:1990rz,Weinberg:1991um} needs to be modified in order to correctly renormalize the theory and a contact interaction has to be included in the Chiral Lagrangian at leading order. From this Chiral Lagrangian, one can then derive the LNV operator between nucleons which can serve as the starting point for many-body calculations. 
We start by briefly reviewing the set of operators at the scale $\Lambda$.

\section{Lepton-number violation in the SM-EFT}
LNV interactions arise at odd dimensions within the SM-EFT \cite{Kobach:2016ami} so that the relevant part of the Lagrangian, at the scale $ \Lambda$ can be written as
\bea
\vL = \vL_{SM}+\vL_{\Delta L=2}^{(5)}+\vL_{\Delta L=2}^{(7)}+\vL_{\Delta L=2}^{(9)}\dots
\eea
where the dots stand for operators beyond dimension-nine. 
Although the dimension-seven (-nine) operators are suppressed by $1/\Lambda^2$ ($1/\Lambda^4$), there are several BSM scenarios, such as the left-right model \cite{Pati:1974yy,Mohapatra:1974hk,Senjanovic:1975rk}, where all of them can play a role. 
Here we will focus on the dimension-five operator and only briefly discuss the effects of the higher-dimensional operators at the scale of a few GeV.
At dimension five there is only one operator~\cite{Weinberg:1979sa} which can be written as,
\bea
\vL_{\Delta L=2}^{(5)} = \ep_{kl}\ep_{mn}(L_k^T\,\mathcal C^{(5)}\,CL_m )H_l H_n,\label{eq:dim3}
\eea
where $C$ is the charge-conjugation matrix, $L$ and $H = 1/\sqrt{2}(0,\, v+h)^T$ are the lepton and Higgs doublets (in unitary gauge), and $v\simeq 246$ GeV is the Higgs vacuum expectation value. The relevant coupling for \NLDBD\ is denoted as $m_{\bt\bt}=-v^2 \mathcal C^{(5)}_{ee}$.
The complete set of dimension-seven operators is known and includes $12$ LNV interactions~\cite{Lehman:2014jma,Babu:2001ex,Bell:2006wi,Liao:2016hru,Liao:2016qyd}, while all dimension-nine operators involving four quarks and two leptons have been classified~\cite{Prezeau:2003xn,Graesser:2016bpz}, but no complete  basis is known. 

The QCD (electroweak) evolution of these operators is known for the dimension-nine (-seven) terms~\cite{Buras:2000if,Buras:2001ra,Cirigliano:2018yza,Liao:2019tep}, allowing one to evolve them from the scale $\Lambda$ to the electroweak scale, where, after integrating out the heavy SM fields, one matches onto a second EFT. This changes the dimension of several operators, so that at scales of a few GeV the Lagrangian involves terms of dimension three, six, seven, and nine.
At dimension three the relevant interaction consists of a Majorana mass for the neutrinos
\bea
\mathcal L_{\Delta L=2}^{(3)\prime }   = - \frac{1}{2} (m_{\nu})_{ij} \, \nu^{T}_{L,\, i} C \nu_{L,\, j}  + \ldots
\eea
where $(m_\nu)_{ij}=-v^2 \mathcal C^{(5)}_{ij}+\dots$, and the dots stand for contributions from operators of dimension-seven and higher.
At dimension six there appear semileptonic four-fermion interactions \cite{Cirigliano:2017djv}
\bea
\mathcal L^{(6)\prime}_{\Delta L = 2}& =& \frac{2 G_F}{\sqrt{2}} \Bigg(
C^{(6)}_{\textrm{VL},ij} \,  \bar u_L \gamma^\mu d_L \, \bar e_{R,i} \,  \gamma_\mu \, C\bar \nu^T_{L,j} + 
C^{(6)}_{\textrm{VR},ij} \,  \bar u_R \gamma^\mu d_R \, \bar e_{R,i}\,  \gamma_\mu  \,C\bar\nu_{L,j}^T \label{lowenergy6}   \\
&  +& \!\!\!\!
C^{(6)}_{ \textrm{SR},ij} \,  \bar u_L  d_R \, \bar e_{L,i}\, C  \bar \nu^T_{L,j} + 
C^{(6)}_{ \textrm{SL},ij} \,  \bar u_R  d_L \, \bar e_{L,i} \, C  \bar\nu_{L,j}^T + 
C^{(6)}_{ \textrm{T},ij} \,  \bar u_L \sigma^{\mu\nu} d_R \, \bar e_{L,i}  \sigma_{\mu\nu}  \, C\bar\nu_{L,j}^T
\Bigg)  +{\rm h.c.}\nn
\eea
These dimension-six terms are generated  by the operators in $\vL^{(7)}_{\Delta L=2}$, which also contribute to the following dimension-seven terms at low energies,
\bea
\mathcal L^{(7)\prime}_{\Delta L = 2} &=& \frac{2 G_F}{\sqrt{2} v} \Bigg( 
C^{(7)}_{\textrm{VL},ij} \,  \bar u_L \gamma^\mu d_L \, \bar e_{L,i} \, C \,  i \overleftrightarrow{\partial}_\mu \bar \nu_{L,j}^T  +
C^{(7)}_{\textrm{VR},ij} \,  \bar u_R \gamma^\mu d_R \, \bar e_{L,i} \, C i \overleftrightarrow{\partial}_\mu \bar \nu^T_{L,j}  \Bigg)  +{\rm h.c.}\label{lowenergy7}
\eea
Due to their origins, one has $C^{(6,7)}_i = \mathcal O(v^3/\Lambda^3)$.  
Finally,  dimension-nine operators~\cite{Graesser:2016bpz,Prezeau:2003xn} with two electrons and four quarks are induced by $\vL^{(7)}_{\Delta L=2}$ and $\vL^{(9)}_{\Delta L=2}$
\bea \label{eq:Lag}
\vL^{(9)\prime}_{\Delta L =2} = \frac{1}{v^5}\sum_i\bigg[\left( C^{(9)}_{i\, \rm R}\, \bar e_R C \bar e^T_{R} + C^{(9)}_{i\, \rm L}\, \bar e_L C \bar e^T_{L} \right)  \, O_i +  C^{(9)}_i\bar e\g_\mu\g_5  C \bar e^T\, O_i^\mu\bigg],
\eea
where $O_i$ and  $O_i^\mu$ are four-quark operators that are Lorentz scalars and vectors, respectively. Their definitions can be found in Ref.~\cite{Cirigliano:2018yza}. 

\section{Chiral  Effective Theory}
We start with the Chiral Lagrangian induced by Eq.\ \eqref{eq:dim3}. In Weinberg's power counting, the leading-order Chiral Lagrangian involves the Majorana neutrino mass, $m_{\bt\bt}$, as well as the one-body weak currents. The latter arises from a single vertex or from pion exchange between the lepton and nucleon line, leading to the amplitude
\bea
\mathcal A^{n\to pe^- \nu} =-\sqrt{2}V_{ud}G_F \bar N \tau^+ \left[v^\mu-2g_A \left(S^\mu + \frac{q^\mu S\cdot q}{\vec q^2+m_\pi^2}\right)\right] N\,\bar e_L\gamma_\mu \nu_L\,,
\eea
where $g_A\simeq 1.27$ is the nucleon axial coupling, $q$ is the momentum transfer, $v^\mu$ and $S^\mu$ are the nucleon velocity and its spin, $V_{ud}$ is a CKM element, and $G_F$ is the Fermi constant.
Combining two insertions of this weak current with the Majorana mass then gives rise to the following LNV Hamiltonian, $H_{\rm LNV} =  2 G_F^2  V_{ud}^2 \  m_{\beta \beta}  
\  \bar e_{L} C \bar e_{L}^T   \ V_\nu  \,,$ with the two-body potential given by
\begin{eqnarray}\label{eq:longrange}
V_{\nu}(\vec q) &=& \tau^{(1) +} \tau^{(2) +} \frac{1}{\vec q^2}  \Bigg\{ 1 - g_A^2 \boldsigma^{(1)} \cdot \boldsigma^{(2)}
+
g_A^2 \, \boldsigma^{(1)} \cdot \vec q\, \boldsigma^{(2)}\cdot \vec q  \    
\frac{2 m_\pi^2  + \vec{q}^2}{(\vec q^2 + m_\pi^2)^2} \Bigg\}~, 
\end{eqnarray}
where $\tau$ and $\sigma$ are isospin and spin matrices, respectively.
This agrees with the commonly employed neutrino potential~\cite{Bilenky:2014uka,Engel:2016xgb} at leading order~\footnote{Within Chiral EFT, dependence on the intermediate nuclear states appears at next-to-next-to-leading order~\cite{Cirigliano:2017tvr}.}. 

The above potential is commonly used as the starting point of \NLDBD\ calculations. These involve the evaluation of phase space integrals over the lepton momenta as well as the matrix element of $V_\nu$ between initial and final nuclear states. While the former are well known~\cite{Horoi:2017gmj}, the latter rely on complicated many-body calculations whose results vary by a factor of two to three between different methods~\cite{Hyvarinen:2015bda,Menendez:2017fdf,Barea:2015kwa,Horoi:2017gmj}. Apart from these theoretical uncertainties, there is the issue that Eq.\ \eqref{eq:longrange} is based on Weinberg's power-counting, which is known to break down in nucleon-nucleon scattering~\cite{Kaplan:1996xu,Beane:2001bc,Nogga:2005hy,Long:2012ve}, making it important to see whether this power-counting scheme is justified in \NLDBD.

\begin{figure}
	\includegraphics[width=0.5\textwidth]{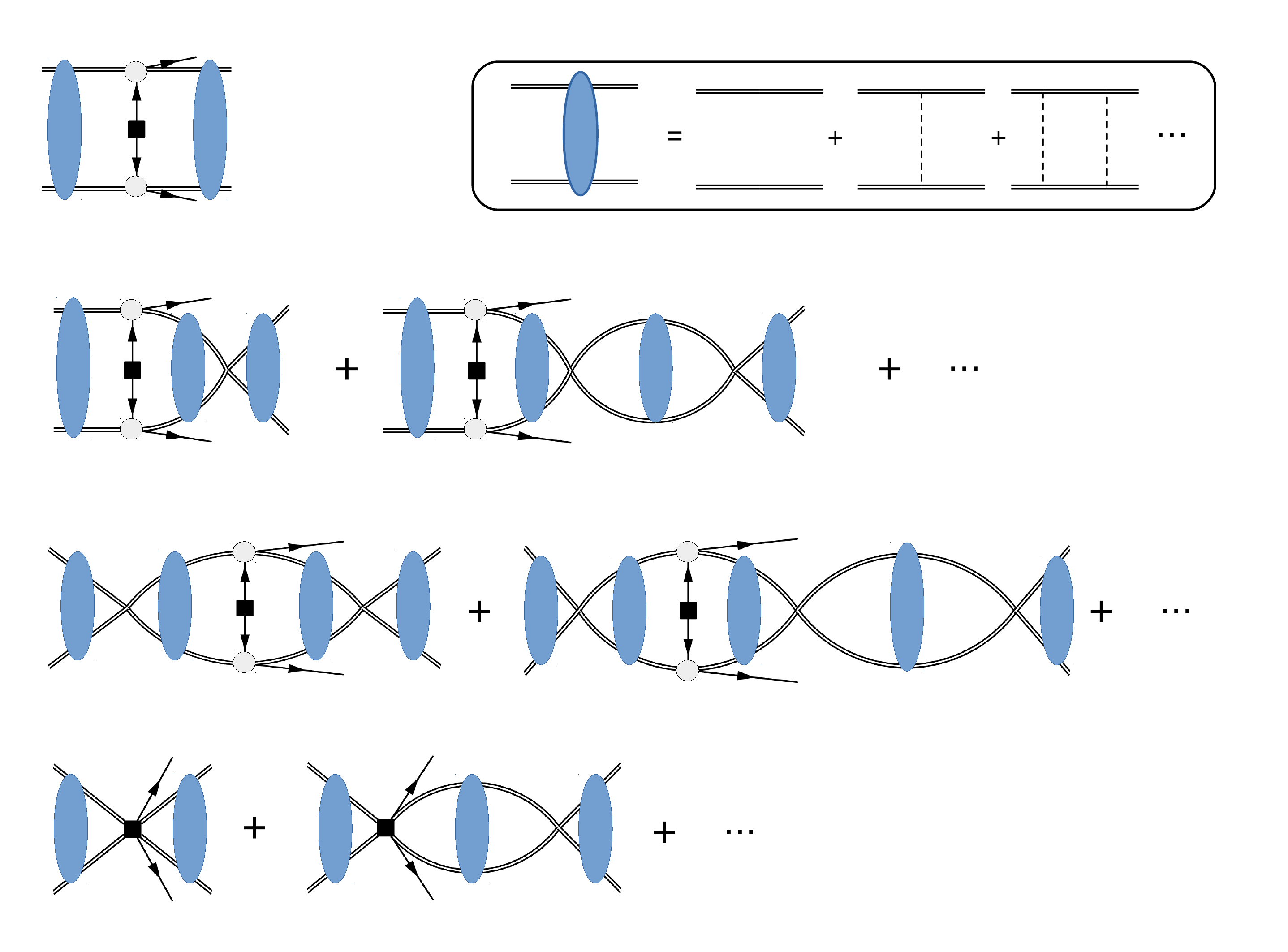}
	\includegraphics[width=0.5\textwidth]{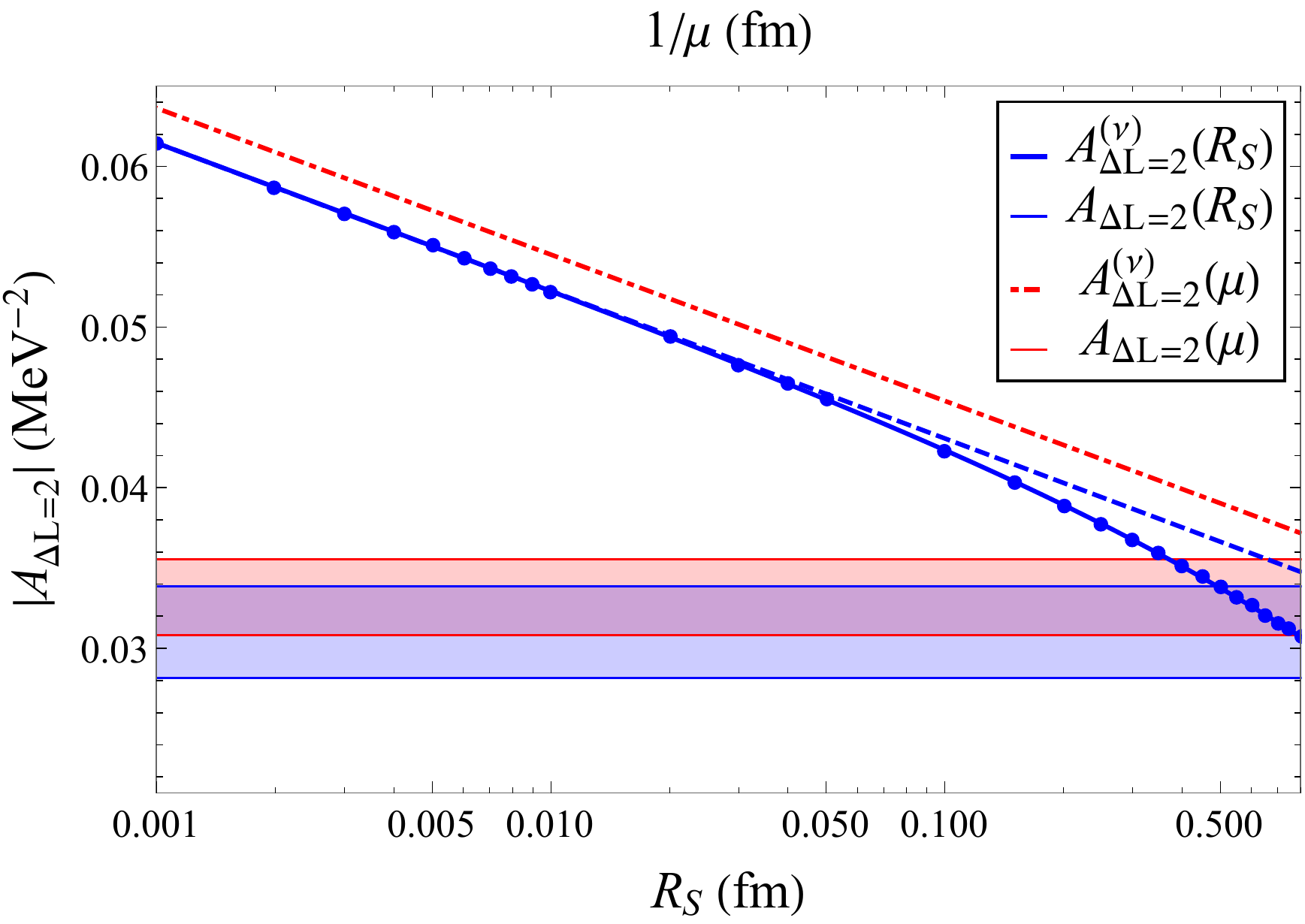}
	\caption{Left panel: Contributions to 
		$n n \rightarrow p p ee$. Double, dashed, and plain lines denote nucleons, 
		pions, and leptons, respectively. 
		Gray circles denote the weak current, and 
		the black square an insertion of  $m_{\beta \beta}$ (or $g_\nu^{NN}$ in the case of the fourth line). Right panel: The $n n \rightarrow p p ee$ amplitude as a function of the regulator. The \MS scheme is shown in red, while the cut-off scheme, explained in the text, is shown in blue.
	}\label{Fig1}
\end{figure}

\subsection{The need for a contact interaction at leading order}
To investigate the power counting, we calculate the simplest possible LNV amplitude, namely, \nnpp, and check whether it can be renormalized. To do so, one needs to dress the LNV potential in Eq.\ \eqref{eq:longrange} with strong interactions. At leading order these consist of pion exchange as well as a contact interaction, which induce the following potential in the $^{1}S_0$ channel
\bea
V_0(\vec q) = \tilde{C} + V_{\pi}(\vec q) \, ,
\quad  
V_{\pi}(\vec q) = -
\frac{g_A^2}{4 F_\pi^2}\frac{m_\pi^2}{\vec q^2 + m_\pi^2}\,,
\eea
where $\tilde C = \Or(F_\pi^{-2})$ parametrizes the short-distance component of the strong interactions. By treating the above potential non-perturbatively, and fitting to the $NN$ scattering length, the strong interactions can be consistently renormalized.

Dressing the LNV potential with the strong potential leads to several classes of diagrams, depicted in the left-hand panel of Fig.\ \ref{Fig1}. The first line shows $V_\nu$ combined with pion exchanges, while in the second line iterations of $\tilde C$ have been added on one side of $V_\nu$. One can show that both types of diagrams give rise to finite results. However, the class of diagrams involving $\tilde C$ inserted on both sides of $V_\nu$ (third line of the figure) leads to a divergence, both in the \MS scheme as well as when one regulates the contact interaction with a Gaussian, $\tilde C \dt^{(3)}(\vec r) \to \frac{\tilde C(R_S)}{(\sqrt{\pi} R_S)^3} e^{-r^2/R_S^2}$. The numerical results for the amplitude in both schemes are shown in the right-hand panel of Fig.\ \ref{Fig1} as a function of the regulator (either $\mu$ in the \MS scheme or $R_S$ in the cut-off scheme). Both schemes exhibit a clear regulator dependence, implying that the obtained results are not properly renormalized.

The regulator dependence can be removed by introducing a LNV contact interaction
\begin{equation}\label{eq:VCT}
V_{\nu, CT} =  -  2 g_\nu^{N\!N}  \  \tau^{(1) +} \tau^{(2) +}   \,,
\end{equation}
where $g_\nu^{NN}$ is an LEC. This potential can be dressed with the strong interactions in the same way as was done for $V_\nu$ (see the last line of the left panel in Fig.\ \ref{Fig1}),  showing that  $g_\nu^{NN}$ can indeed absorb the regulator dependence. 
Within the \MS scheme, this LEC then follows the renormalization-group equation,
\be
\mu \frac{d}{d\mu}\left[\left(\frac{4\pi}{m_N \tilde C}\right)\sq\tilde{g}_\nu^{N\!N}\right] = \frac{1}{2} \left(1 + 2 g_A^2\right)~,
\ee
 which suggests that the combination in square brackets should be $\Or(1)$, implying $g_\nu^{NN}=\Or(F_\pi^{-2})$, in contrast to Weinberg's power counting, $g_\nu^{NN}=\Or\left((4\pi F_\pi)^{-2}\right)$.
 
As the finite part of $g_\nu^{NN}$ is unknown, it is hard to quantify its impact on \NLDBD\ calculations. Preferably, one would determine it from a lattice QCD calculation of \nnpp. At present, however, only order-of-magnitude estimates are available based on Chiral symmetry, which relates $g_\nu^{NN}$ to contact interactions induced by the exchange of hard photons~\cite{Cirigliano:2018hja}. 
Using this estimate to compute the LNV amplitude induced by $V_{\nu}+V_{\nu,CT}$ leads to the horizontal lines in Fig.\ \eqref{Fig1}, showing that the amplitude can indeed be made regulator independent and that the impact of $g^{NN}_\nu$ is at the $10\%$ level for $R_S = 0.6$ fm in \nnpp. In the case of $^{12}$Be$\to ^{12}$C, which is closer to the transitions of experimental interest as it is changes the total isospin~\cite{Pastore:2017ofx}, the impact is at the $\sim 60\%$ level~\cite{Cirigliano:2018hja}.

\begin{figure}
	\includegraphics[trim={.7cm 0 0 .5cm},clip,width=0.5\textwidth]{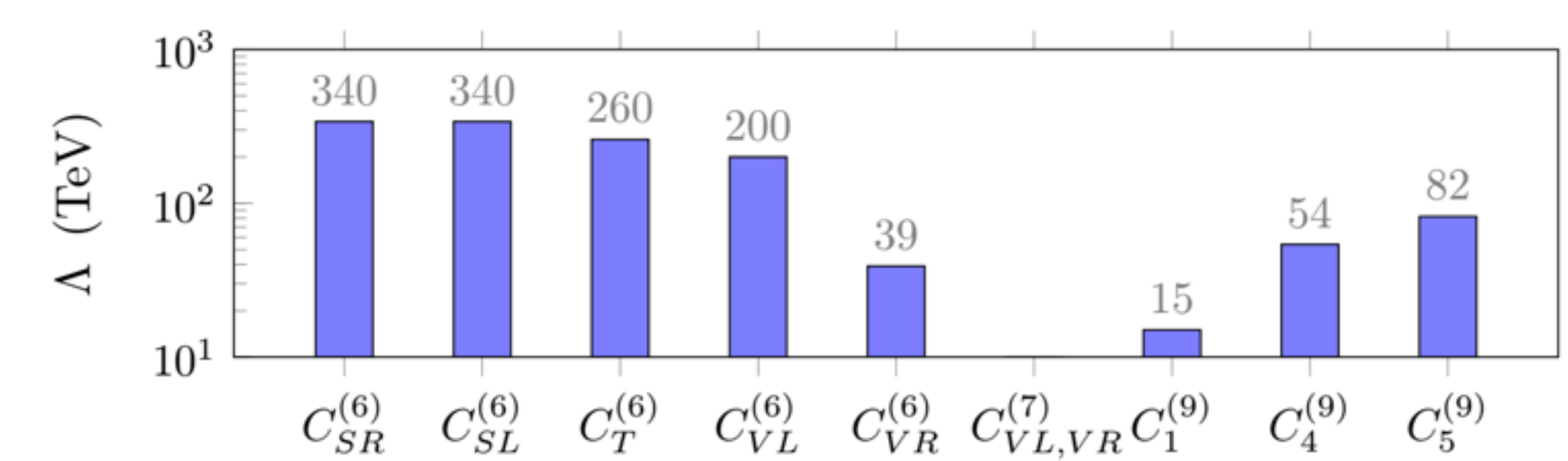}
	\includegraphics[trim={-.3cm 0 0 0},width=0.5\textwidth]{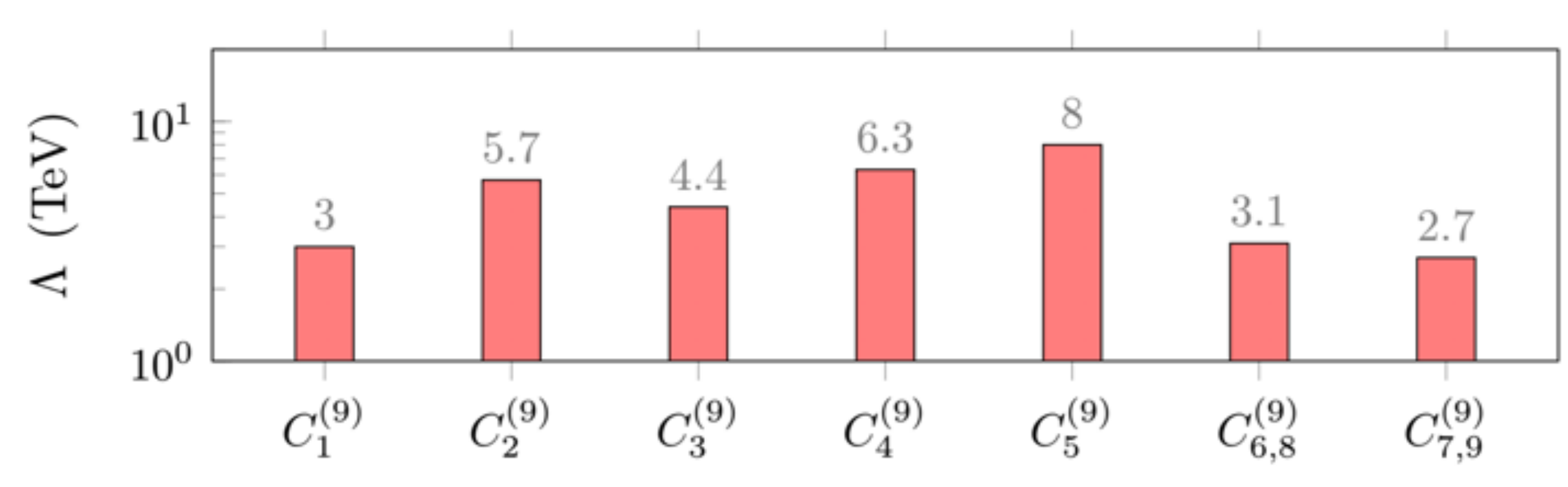}
	\caption{Limits on the Wilson coefficients in Eqs.\ \eqref{lowenergy6}, \eqref{lowenergy7}, and \eqref{eq:Lag}. The left panel depicts the limits on the couplings generated at dimension seven and assume $C_i = v^3/\Lambda^3$, while the right panel shows the constraints on couplings induced by dimension-nine operators, assuming $C_i = v^5/\Lambda^5$.
	}\label{Fig:limits}
\end{figure}

\subsection{Dimension-seven and -nine contributions}
The  matching to Chiral EFT can be repeated for the operators in Eqs.\ \eqref{lowenergy6}, \eqref{lowenergy7}, and \eqref{eq:Lag}. Within Weinberg's power counting the scalar dimension-nine operators then mainly induce $\pi\pi \bar ee^c$ interactions, while the vector operators generate $\pi \bar p n \bar e e^c$ and $(\bar pn) (\bar pn) \bar ee^c$ terms. Instead, the operators in Eqs.\ \eqref{lowenergy6} and \eqref{lowenergy7} would mainly induce one-body interactions, $ \bar p n \bar e\nu^c$. The corresponding LECs have been calculated on the lattice for the scalar dimension-nine terms~\cite{Nicholson:2018mwc},  most of those needed for $C_{i}^{(6,7)}$ are the known nucleon charges~\cite{Cirigliano:2017djv}, while the LECs for the dimension-nine vector terms are currently unknown. Furthermore, the issue with Weinberg's power counting as described above reappears for several of the higher-dimensional operators. Similar arguments then suggest that one needs additional contact interactions for the scalar dimension-nine operators, and some of the $C_{i}^{(6,7)}$. On the other hand, the needed nuclear matrix elements have all been evaluated in the literature. Using these~\cite{Menendez:2017fdf}, together with some assumptions on the LECs~\cite{Cirigliano:2018yza} and the experimental limit~\cite{KamLAND-Zen:2016pfg}, then allows one to set the constraints depicted in Fig.\ \ref{Fig:limits}.
\newline
\newline
We conclude that EFTs provide a systematic way to assess the contributions of LNV sources to \NLDBD. Despite large nuclear and hadronic uncertainties, this typically leads to stringent limits in the case of dimension-seven operators, while dimension-nine terms are constrained to be at the TeV scale, see Fig.\ \ref{Fig:limits}.  
In addition, consistent renormalization requires one to include a new short-range interaction at leading order, which already affects the light Majorana-neutrino exchange mechanism.

\section*{Acknowledgments}

I would like to thank the organizers for the opportunity to present this work at the electroweak session of Recontres de Moriond, and for an interesting and enjoyable meeting.  This work was supported in part by DOE Grant No.\ DE-SC0009919.

\section*{References}

\bibliography{bibliography}
\end{document}